# Fast Algorithm for Finding Maximum Distance with Space Subdivision in $E^2$

Vaclav Skala[1], Zuzana Majdisova[1]

[1] Faculty of Applied Sciences, University of West Bohemia,
Univerzitni 8, CZ 30614 Plzen, Czech Republic

**Abstract.** Finding an exact maximum distance of two points in the given set is a fundamental computational problem which is solved in many applications. This paper presents a fast, simple to implement and robust algorithm for finding this maximum distance of two points in $E^2$. This algorithm is based on a polar subdivision followed by division of remaining points into uniform grid. The main idea of the algorithm is to eliminate as many input points as possible before finding the maximum distance. The proposed algorithm gives the significant speed up compared to the standard algorithm.

## 1 Introduction

Finding a maximum distance of two points in the given data set is a fundamental computational problem. The solution of this problem is needed in many applications. A standard brute force (BF) algorithm with $O(N^2)$ complexity is usually used, where $N$ is a number of points in the input dataset. If large sets of points have to be processed, then the BF algorithm leads to very bad time performance. Typical size of datasets in computer graphics is usually $10^5$ and more points. Therefore the processing time of the BF algorithm for such sets is unacceptable.

However, our main goal is to find the maximum distance, not all the pairs of two points having a maximum distance. Therefore the complexity of this algorithm should be lower.

Various approaches, how to solve finding the maximum distance, are described in [9]. Other algorithms for finding the maximum distance of two points are in [1], [7].

### 1.1 Brute Force Algorithm

The standard BF algorithm for finding a maximum distance in set of points uses two nested loops. We can find such type of algorithms in many books dealing with fundamental algorithms and data structures, e.g. [4], [6]. In general, the BF algorithm can be described by Algorithm 1.

```
//Square of the distance
FUNCTION distance(A,B: point)
  distance := (A.x  - B.x)^2 + (A.y - B.y)^2;
END FUNCTION

dist := 0;
FOR i := 1 to N-1 do
  FOR j := i + 1 to N do
    dij := distance(Xi, Xj);
    IF dist < dij THEN
      dist := dij;
    END IF
  END FOR
END FOR
dist := SQRT(dist);
```

**Algorithm 1.** Brute force algorithm

Complexity of Algorithm 1 is clearly $O(N^2)$ and thus run time significantly increases with size of the input dataset.

In practice, we can expect that points in input set are not organized in a very specific manner and points are more or less uniformly distributed. In this case, we can use "output sensitive" algorithms which lead to efficient solutions. We propose such algorithm in Section 2.

## 2 Proposed Algorithm

In this section, we introduce a new algorithm for finding a maximum distance of two points in the given dataset in $E^2$. The main idea of this algorithm is to eliminate as many input points as possible using an algorithm with $O(N)$ complexity using space subdivision and determines the maximum distance for the remaining points with $O(k^2)$ complexity, where $k \ll N$. We use polar space subdivision for this elimination of points.

This section is organized as follows. In Section 2.1, we present the first step of the algorithm which is an axis aligned bounding box (AABB) and an initial convex polygon construction followed by the location of points inside the initial convex polygon. Section 2.2 describes how to divide the points into non-overlapping $2D$ triangular shape sectors. Section 2.3 presents reduction of the points [2] which have absolutely no influence on the value of maximum distance. In Section 2.4, we describe the division of remaining points into uniform $2D$ grid. Finally, the finding of the maximum distance of two points is made in Section 2.5.

## 2.1 Location of Points inside Initial Polygon

An important property is that two points with maximal distance are lying on the convex hull of a given set of points [10]. This fact is apparent if we consider a case in which two points with the largest distance are part of the convex hull. It is then obvious that there are another two points with larger distance. We also know that the most extreme point on any axis is part of the convex hull. These properties are used to significantly speedup the proposed algorithm for finding the exact maximal distance.

At the beginning of our proposed algorithm, we need to find the exact extremal points in both axes, i.e. axis aligned bounding box (AABB) of a given dataset. The time complexity of this step is $O(N)$. So we generally get four distinct extremal points or less.

Now, we can create a convex polygon using these extremal points, see Fig. 1. One important property of this polygon is that any point lying inside has no influence on the value of maximal distance. Due to this fact, we can perform a fast and simple initial test for a point inside/outside the initial polygon and discard a lot of points.

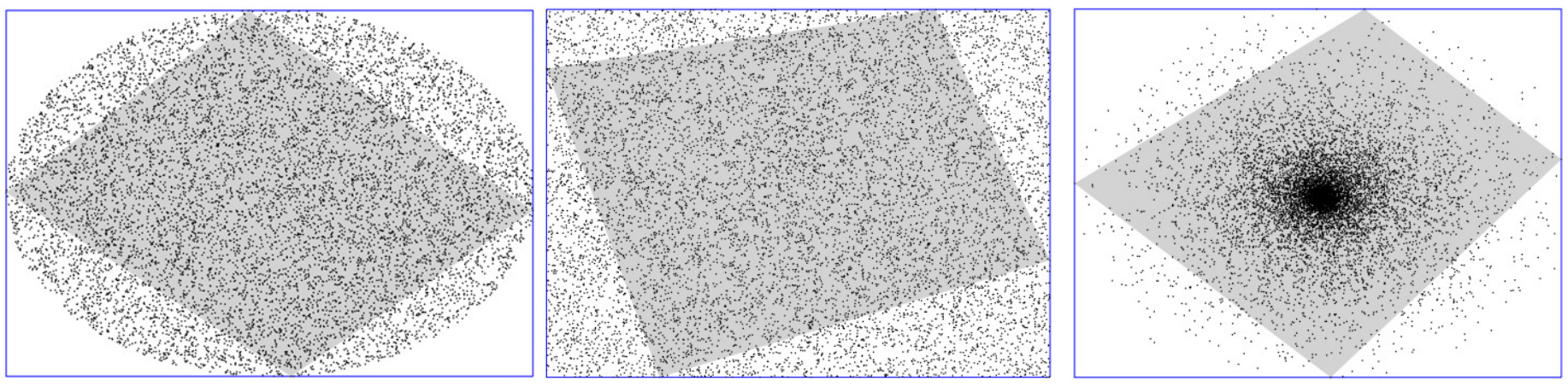

a) b) c)

**Fig. 1.** Location of AABB and initial testing polygon for $10^4$ points: a) uniform points in ellipse, b) uniform points in rectangle, c) Gauss points.

The location test of a point inside a polygon can be performed as follows. Each edge of the polygon is an oriented line and so we can calculate outer product [10]:

$$f_i(\boldsymbol{x}) = \boldsymbol{v}_i \wedge (\boldsymbol{x} - \boldsymbol{x}_i) = \begin{vmatrix} v_{xi} & v_{yi} \\ x - x_i & y - y_i \end{vmatrix}, \tag{1}$$

where $\boldsymbol{x}$ is the point and edge with index $i$ is determined by point $\boldsymbol{x}_i$ and direction vector $\boldsymbol{v}_i = (v_{xi}, v_{yi})$. If the polygon has an anticlockwise orientation and outer product $f_i(\boldsymbol{x}) \leq 0$ for at least one $i \in \{0,1,2,3\}$, then point $\boldsymbol{x}$ does not lie inside the polygon and has to be further processed. Otherwise, point $\boldsymbol{x}$ can be discarded as it is inside.

## 2.2 Division of Points into Polar Sectors

Only the points which lie outside or on the boundary of the initial convex polygon will be further processed. Firstly, we perform the division of AABB into eight non-overlapping $2D$ triangular shape sectors, i.e. polar subdivision. This division of AABB is using a center point and angular division, see Fig. 2. The center point $\boldsymbol{C}$ is determined as the average of all corners of the AABB.

When we do the division of points into non-overlapping sectors, we also determine angle between the $x$-axis and the vector $\boldsymbol{s} = \boldsymbol{x} - \boldsymbol{C}$ for each point $\boldsymbol{x}$. This can be performed using two different calculations. One way is to use an exact angle from 0 to $2\pi$. For this approach, we have to calculate the angle using the following formula:

$$\varphi_e = \mathrm{arctg2}\left(\boldsymbol{v}_x, \boldsymbol{v}_y\right). \tag{2}$$

However, calculation of function arctg2 takes a lot of computing time. Therefore, we use a simplified calculation of approximated angle. When the angle is determined, we have to locate the exact sectors (half of the quadrant for square AABB), where the point is located, and then calculate the intersection with the given edge. Calculation of the intersection with the given edge of AABB is easy. The distribution of simplified angle can be seen in Fig. 2. Calculation of simplified angle is faster than the formula (2).

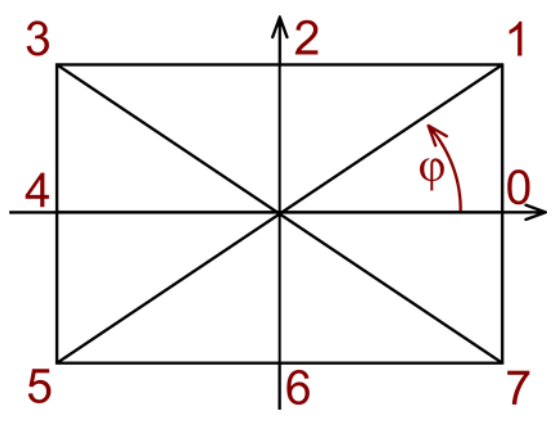

**Fig. 2.** Non-overlapping sectors for division and uniform distribution of simplified angle on AABB. Angle $\varphi \in [0,8)$ instead of $[0,2\pi)$.

Now we have the procedure how to calculate the simplified angle and therefore we are able to divide the points into sectors to which the given points belong.

For each sector with index $i$, one minimal point $\boldsymbol{R}_i^{min}$ is determined. This point has the minimum (from all points in a sector) distance from the nearest corner of AABB. (Note that the nearest corner of AABB lies in the same quadrant as the point.) The initial points $\boldsymbol{R}_i^{min}$ are lying on the edges of the initial polygon, see Fig. 3. These points can be calculated as an intersection point of the middle axis of a sector and the edge of the initial polygon.

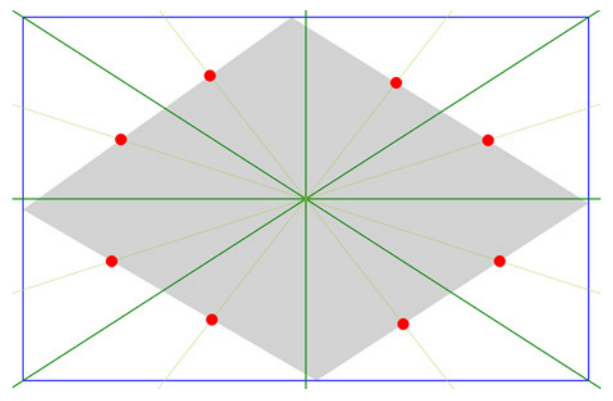

**Fig. 3.** Visualization of initial $R_i^{min}$ points (red dots on the edges of the initial polygon).

All minimum points $\boldsymbol{R}_i^{min}$ are connected into a polygon with vertices $\boldsymbol{R}_1^{min}, \ldots, \boldsymbol{R}_8^{min}$.

For each new point we have to check whether the distance from this point to the nearest corner of AABB is smaller than the distance from $\boldsymbol{R}_i^{min}$ to the same corner of AABB. If this is true, then we have to replace point $\boldsymbol{R}_i^{min}$ with a processed point, add this point into the sector with index $i$ and recalculate the test lines $l^-$ and $l^+$, see Fig. 4. Otherwise we continue with the next step.

In the next step, we check whether the processed point lies over or under the test line segments $l^-$ and $l^+$. We can compare the angle of the point with the angle of point $\boldsymbol{R}_i^{min}$. If the angle is smaller, then we have to use the line $l^-$, otherwise we have to use the line $l^+$. If the point lies under the test line, it can be eliminated, because such a point has no influence on the value of maximum distance. Otherwise we add this point into the sector with index $i$.

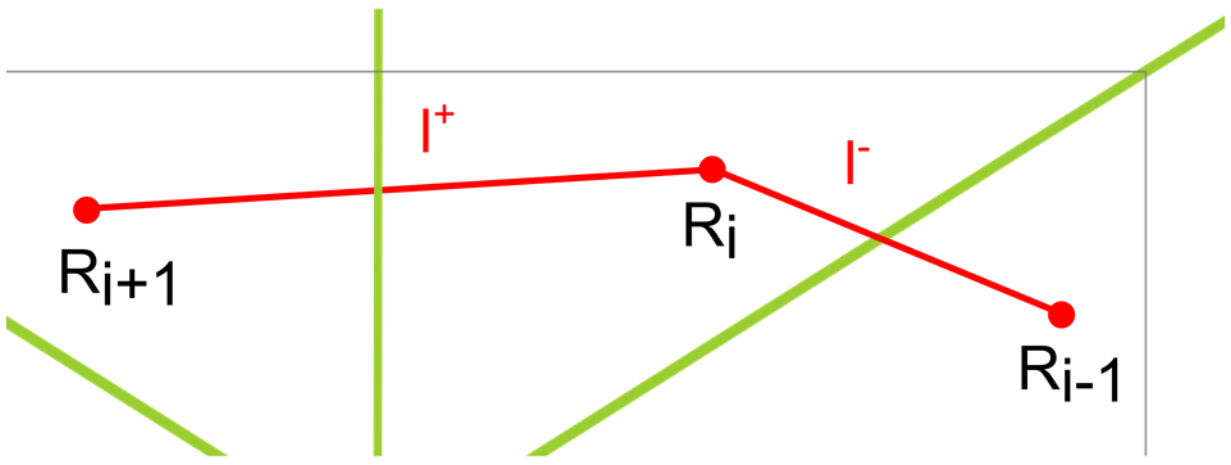

**Fig. 4.** Visualization of test lines $l^-$ and $l^+$.

### 2.3 Reduction of Points for Testing

All points, which can have an influence on the value of maximum distance, are already divided into polar sectors. We gave points $\boldsymbol{R}_i^{min}$ some initial values before starting to divide the points into non-overlapping sectors and we used them to check whether to add or eliminate a point. Values of points $\boldsymbol{R}_i^{min}$ have changed during the division process; hence we recheck all remaining points using the final values of points $\boldsymbol{R}_i^{min}$. Moreover, we perform union of the vertices of initial polygon and minimum points $\boldsymbol{R}_i^{min}$ before new testing and connect them into a polygon, see Fig. 5a).

In this step, we check whether the processed point lies over or under the line segments $l^{--}$, $l^-$, $l^+$ and $l^{++}$, see Fig. 5b). We select the concrete test line according to the angle again.

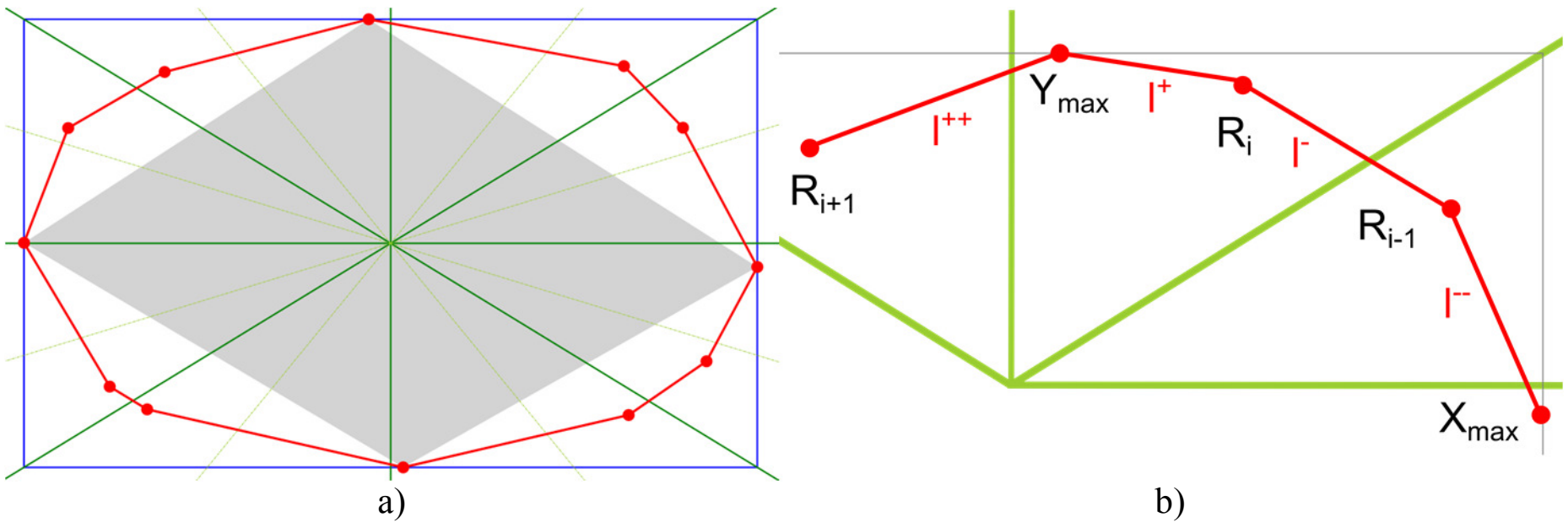

**Fig. 5.** Visualization of test lines for rechecking all remaining points

We minimize the number of points, which have an influence on the largest distance, using this step. Final sets of remaining points for input datasets with different distributions of points are shown in Fig. 6.

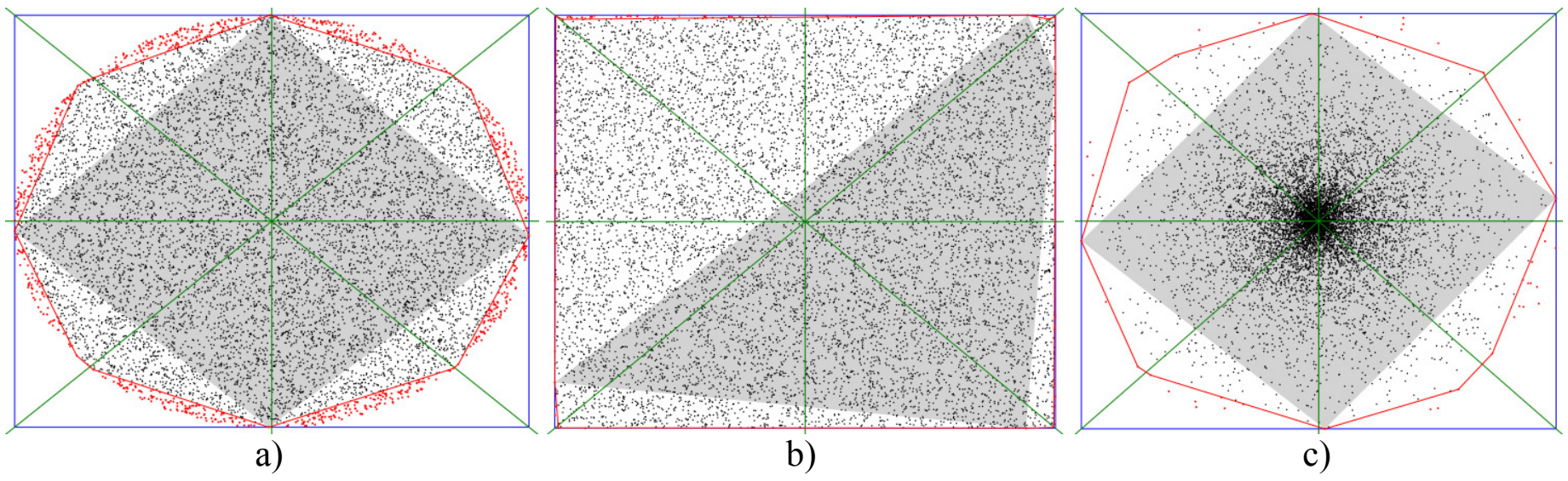

**Fig. 6.** Remaining points (red dots) which have influence on the maximum distance ($10^4$ input points): a) uniform points in ellipse, b) uniform points in rectangle, c) Gauss points.

## 2.4 Division of Remaining Points into Uniform Grid

We have a set of suspicious points, i.e. points which can have an influence on the final maximum distance. In this step, these suspicious points will be further processed. Firstly, we define the uniform grid of AABB. This uniform grid contains $k \times k$ cells. Thus, each cell has index $i = row \cdot k + col$, width $dx$ and height $dy$ where:

$$dx = \frac{AABB_{width}}{k}, dy = \frac{AABB_{height}}{k}. \quad (3)$$

Now, we can perform the division of suspicious points into a defined uniform grid. We are able to calculate the exact index of a cell to which the given point $\boldsymbol{x}$ belongs using following formulae:

$$row = \left\lfloor \frac{y - y_{min}}{dy} \right\rfloor, col = \left\lfloor \frac{x - x_{min}}{dx} \right\rfloor, \quad (4)$$

where $x_{min}$ and $y_{min}$ are the coordinates of bottom left corner of AABB, see Fig. 7.

After performing previous step, we determined all possible pairs of nonempty cells. Moreover, for each pair of nonempty cells, the shortest distance $d_{ij}^{cell}$, i.e. the distance of the nearest corners of cells, and the largest distance $D_{ij}^{cell}$, i.e. the distance of the farthest corners of cells, are determined, see Fig. 7.

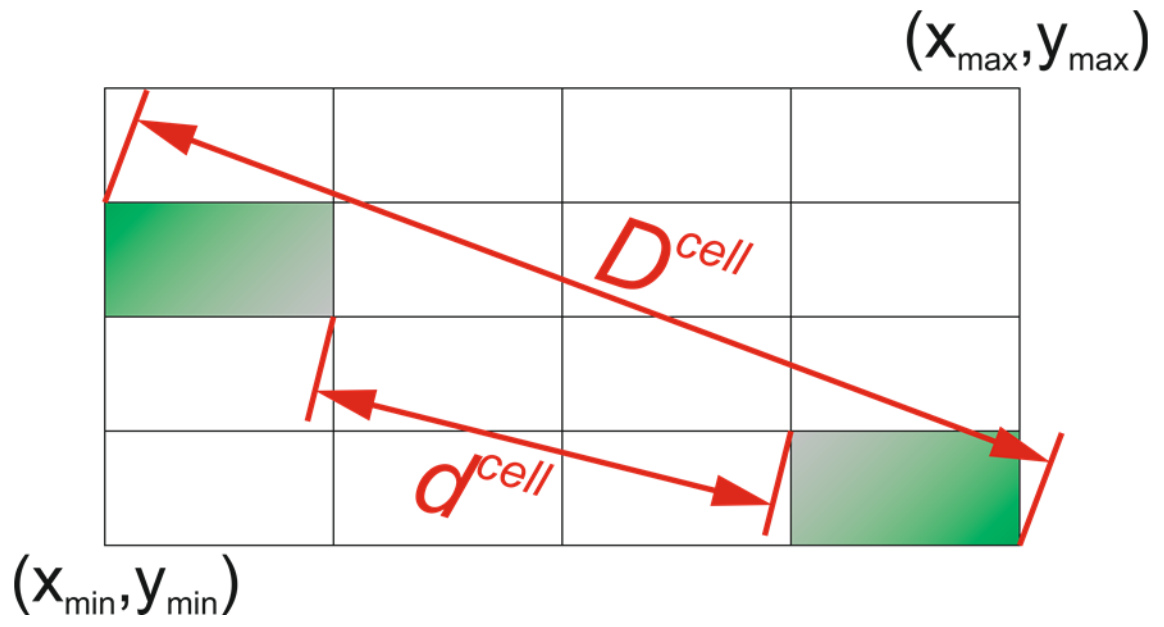

**Fig. 7.** Uniform grid of AABB. Value $D^{cell}$ presents the largest distance of two cells and $d^{cell}$ presents the shortest distance of two cells.

### 2.5 Find Maximal Distance of Two Points

Now a maximum distance of two points in the given dataset can be found by following steps. We determine the maximum value $d_{max}^{cell}$ from the shortest distances $d_{ij}^{cell}$ which were calculated for all pairs of nonempty cells. When this value is known, we can eliminate all pairs of nonempty cells for which the largest distance $D_{ij}^{cell}$ is smaller than $d_{max}^{cell}$.

For remaining pairs of nonempty cells, we perform the following. For each pair of nonempty cells, the maximum distance $D_{ij}$ between points in these cells is determined, i.e. we calculate all distances from points in one cell to points in second cell and determine their maximum. Finally, we find the maximum value of these maximum distances $D_{ij}$.

## 3 Experimental Results

The proposed algorithm has been implemented in C# using .Net Framework 4.5 and tested on datasets using a PC with the following configuration:

- CPU: Intel® Core™ i7-2600 (4 × 3.40 GHz)
- memory: 16 GB RAM
- operating system Microsoft Windows 7 64bits

### 3.1 Distribution of Points

The proposed algorithm for finding the maximum distance of two points has been tested using different datasets. These datasets have different types of distributions of points. For our experiments, we used well-known distributions such as randomly distributed uniform points in an ellipse, uniform points in a rectangle or points with a Gaussian distribution. Other distributions used were Halton points and Gauss ring points. Both of these distributions are described in the following text.

**Halton Points.** Construction of a Halton sequence is based on a deterministic method. This sequence generates well-spaced "draws" points from the interval $[\mathbf{0}, \mathbf{1}]$. The sequence uses a prime number as its base and is constructed based on finer and finer prime-based divisions of sub-intervals of the unit interval. The Halton sequence [3] can be described by the following recurrence formula:

$$Halton(p)_k = \sum_{i=0}^{\lfloor \log_p k \rfloor} \frac{1}{p^{i+1}} \left( \left\lfloor \frac{k}{p^i} \right\rfloor \bmod p \right), \tag{5}$$

where $p$ is the prime number and $k$ is the index of the calculated element.

For the $2D$ space, subsequent prime numbers are used as a base. In our test, we used $\{2,3\}$ for the Halton sequence and we got a following sequence of points in a rectangle:

$$\begin{aligned} Halton(2,3,5) = \Big\{ &\left(\frac{1}{2}a, \frac{1}{3}b\right), \left(\frac{1}{4}a, \frac{2}{3}b\right), \left(\frac{3}{4}a, \frac{1}{9}b\right), \left(\frac{1}{8}a, \frac{4}{9}b\right), \left(\frac{5}{8}a, \frac{7}{9}b\right), \\ &\left(\frac{3}{8}a, \frac{2}{9}b\right), \left(\frac{7}{8}a, \frac{5}{9}b\right), \left(\frac{1}{16}a, \frac{8}{9}b\right), \left(\frac{9}{16}a, \frac{1}{27}b\right), \dots \Big\} \end{aligned} \tag{6}$$

where $a$ is a width of the rectangle and $b$ is a height of the rectangle.

Visualization of the dataset with $10^3$ points of the Halton sequence from (6) can be seen in Fig. 8. We can see that the Halton sequence in $2D$ space covers this space more evenly than randomly distributed uniform points in the same rectangle.

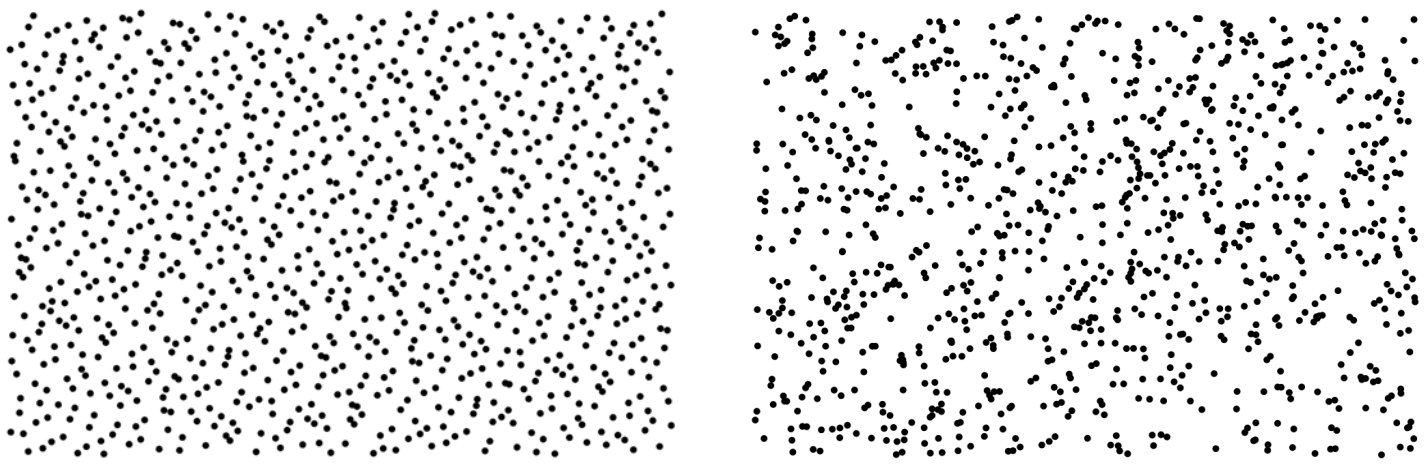

**Fig. 8.** $2D$ Halton points generated by $Halton(2,3)$ (left) and $2D$ random points in a rectangle with uniform distribution (right). Number of points is $10^3$ in both cases.

**Gauss Ring Points.** It is a special distribution of points in $\mathbf{2D}$. Each point is determined as follows:

$$P = \left[a \cdot \varepsilon \cdot \cos\big(rand(0,2\pi)\big), b \cdot \varepsilon \cdot \sin\big(rand(0,2\pi)\big)\right]$$
$$\varepsilon = 0.5 + 0.5 \cdot sign \cdot rand_{Gauss} \tag{7}$$

where $a$ is a length of semi-major axis, $b$ is a length of semi-minor axis, $sign$ is a randomly generated number from set $\{-1,1\}$, $rand_{Gauss}$ is a randomly generated number with Gauss distribution from interval $[0,\infty)$ and $rand(0,2\pi)$ is a random number with uniform distribution from 0 to $2\pi$.

Visualization of the dataset with $10^3$ Gauss ring points can be seen in Fig. 9. We can see that this dataset consists of a large set of points, which are close to the ellipse, and a small set of points, which are far from this ellipse.

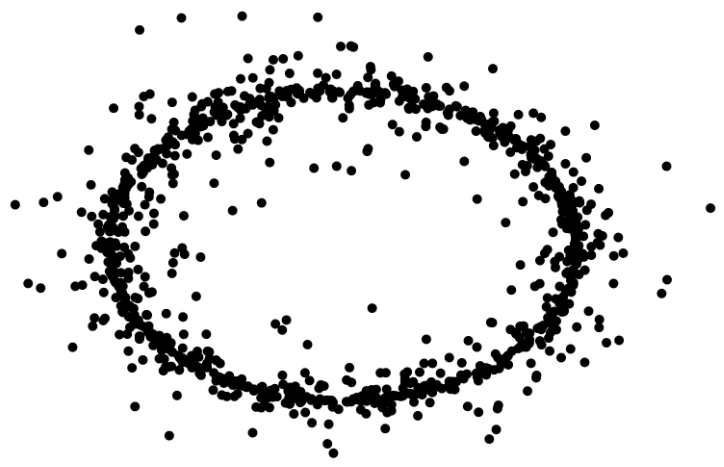

**Fig. 9.** $2D$ Gauss ring points. Number of points is $10^3$.

### 3.2 Optimal Size of Grid

In the proposed approach, the remaining points are divided into uniform grid $k \times k$ after their elimination by polar division. The size of the grid has significantly influence on the number of pairs of points for which their mutual distance is determined. Simultaneously the time complexity is increasing with growing size of the grid. Therefore, we need know an estimation of the optimal size of the grid, which should be dependent on the distribution of points and on the number of points. Therefore, we have to measure it for each type of input points separately.

We measured the time performance of our proposed algorithm for different distributions of points, different numbers of points and different sizes of grid. Measurement for $10^7$ points is presented in Fig. 10. For all tested distributions of input points, we can see that the time performance decreases with the increasing size of grid until the optimal size of the grid is achieved. After that time, the complexity increases with the increasing number of divisions. Moreover, for all tested distributions of input points, except uniform points in the ellipse, it can be seen that the time complexity is practically independent on size of the grid. This is due to the fact that size of set of suspicious points is very small and the number of nonempty cells is small too, see Fig. 6 b)- Fig. 6 c). Thus the time complexity of division into uniform grid and consequent calculation is almost insignificant.

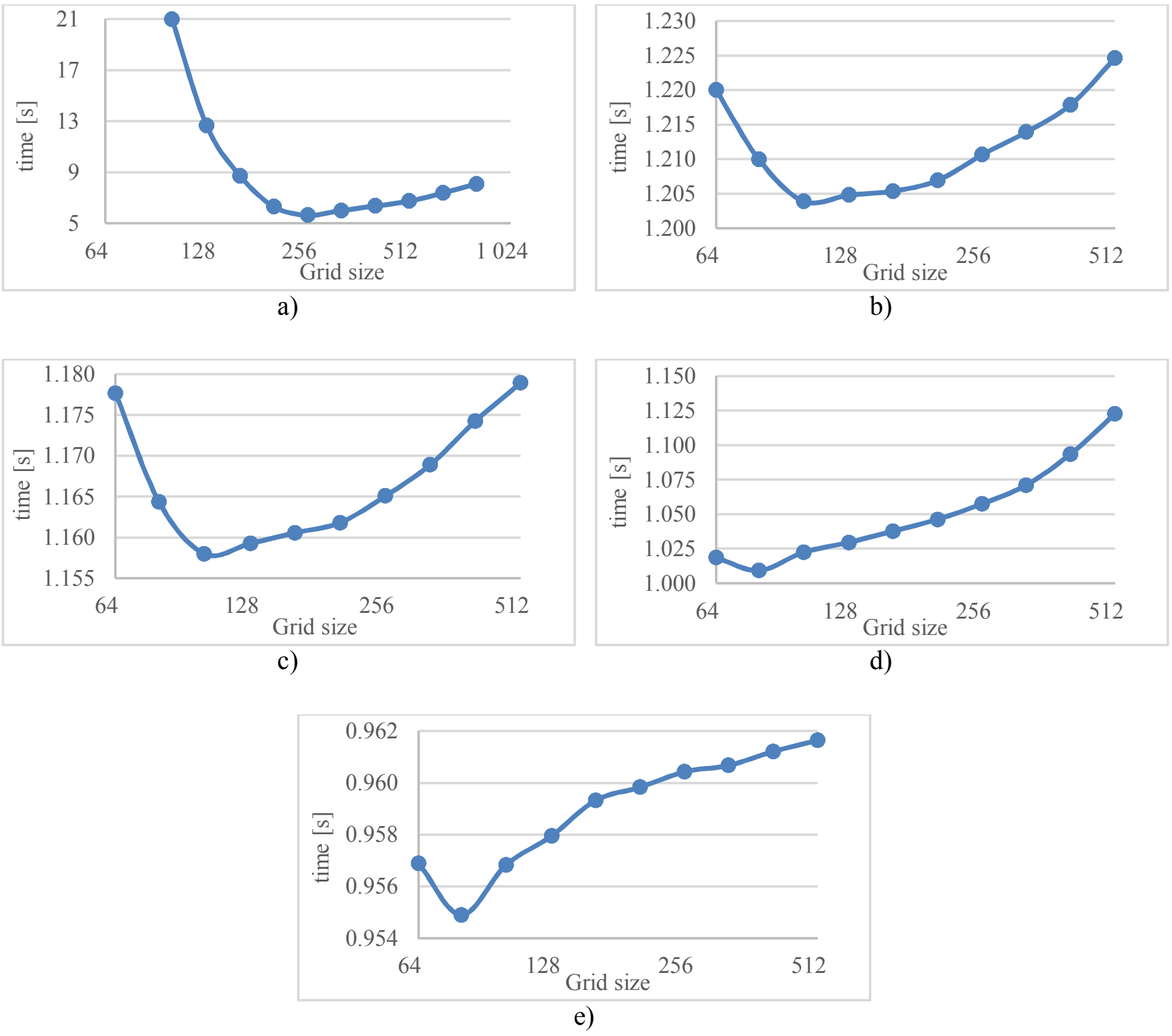

**Fig. 10.** The time performance of algorithm for finding maximum distance of two points for different points distributions and different size of grid. The size of grid denotes the number of cells in one axis. The number of input points is $10^7$. Distribution of points are: a) uniform points in ellipse, b) uniform points in rectangle, c) Halton points, d) Gauss points, e) Gauss ring points.

Fig. 11 presents the optimal size of grid for different distributions of points and different numbers of points. It can be seen that the optimal size of grid increases with the increasing number of points. Moreover, we can see that for uniform distribution of points in the ellipse is needed larger size of the grid than for other tested distributions. This is due to the fact that for this distribution of points is the number of suspicious points substantially larger.

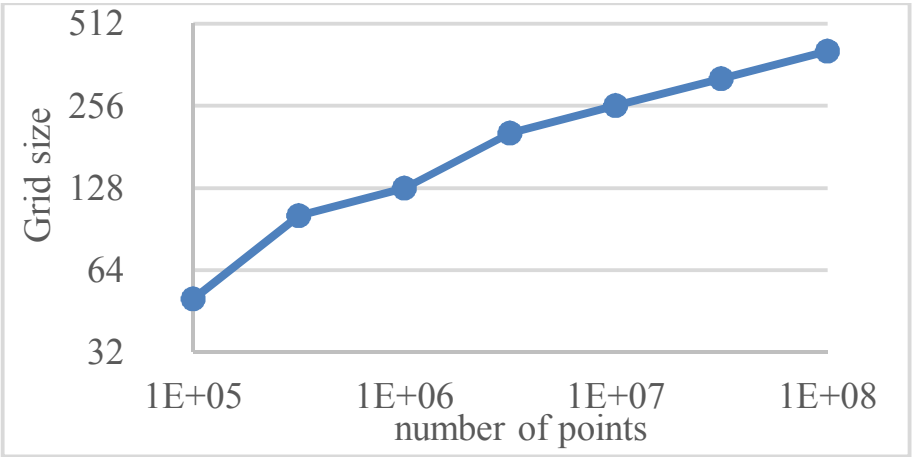

a)

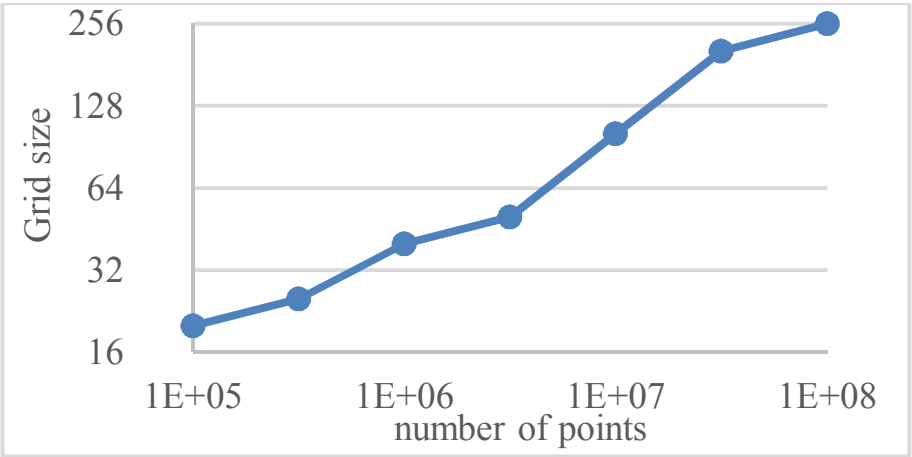

b)

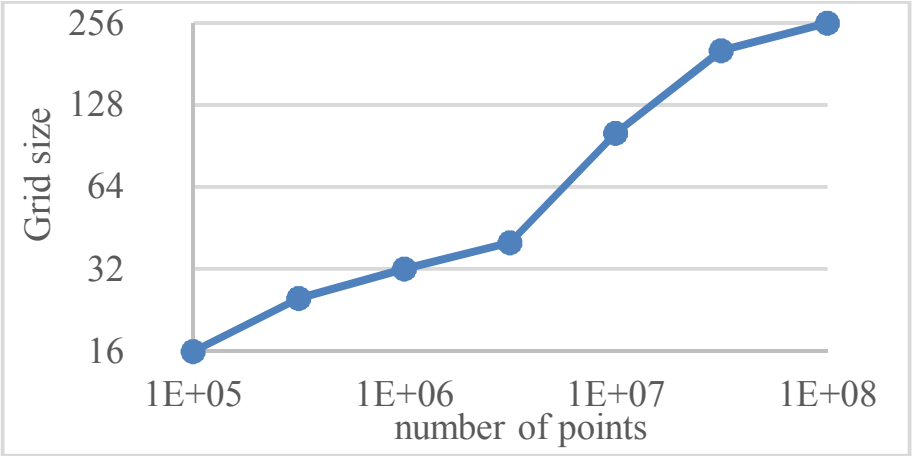

c)

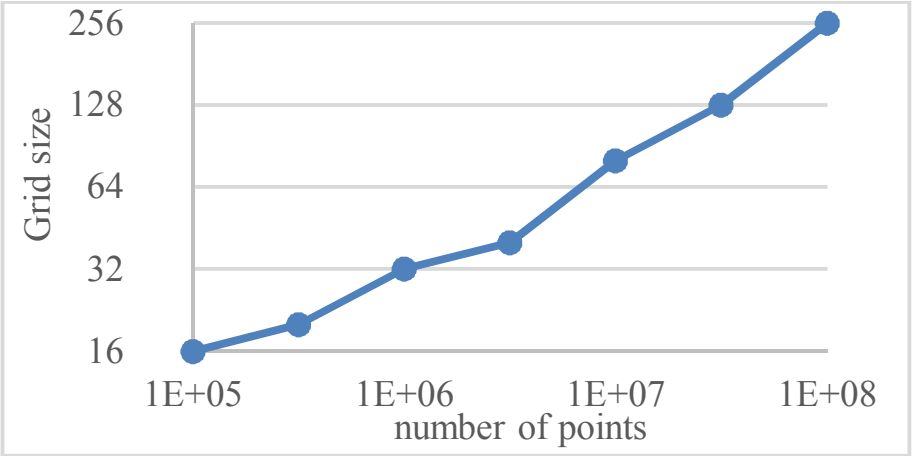

d)

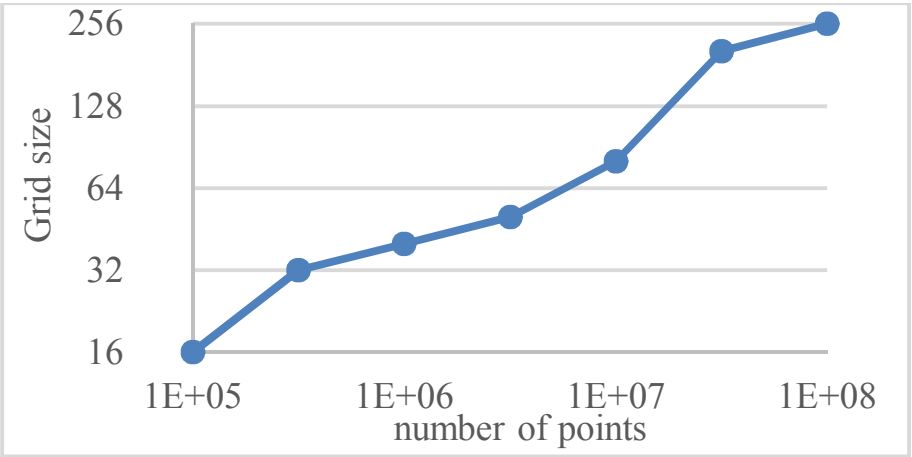

e)

**Fig. 11.** Optimal number of grid size for algorithm for finding maximum distance of two points for different points distributions and different number of points. The size of grid denotes the number of cells in one axis. Distribution of points are: a) uniform points in ellipse, b) uniform points in rectangle, c) Halton points, d) Gauss points, e) Gauss ring points.

Evaluating experimental results for different distributions of points and different numbers of input points, i.e. $10^6$, $\sqrt{10} \cdot 10^6$, $10^7$, $\sqrt{10} \cdot 10^7$ and $10^8$, including results from Fig. 10 and Fig. 11, we came to the following conclusion.

The optimal size of the grid is dependent on the number of input points, more precisely the size of the grid is dependent on number of suspicious point. Size of the grid has to increase with the increasing number of points.

### 3.3 Time Performance

In some applications, the time performance is one of an important criterion. Therefore, running times were measured for different number of input points with different distributions of points. Measurements were performed many times and average running times, calculated from the measured results, are in Table 1. Moreover, we can see these running times in Fig. 12.

**Table 1.** The time performance of convex hull for different number of input points and different distributions of points.

| | Time [ms] | | | | |
|---|---|---|---|---|---|
| **Number of points** | **Uniform ○** | **Uniform □** | **Halton** | **Gauss** | **GaussRing** |
| 1E+5 | 32.9 | 11.5 | 11.0 | 9.0 | 8.8 |
| √10E+5 | 137.6 | 37.4 | 36.3 | 30.6 | 29.8 |
| 1E+6 | 466.2 | 119.1 | 113.5 | 93.3 | 93.4 |
| √10E+6 | 1 745.5 | 367.8 | 355.8 | 315.0 | 296.0 |
| 1E+7 | 5 631.3 | 1 203.9 | 1 158.0 | 1 009.2 | 954.9 |
| √10E+7 | 17 976.5 | 3 596.6 | 3 579.0 | 3 221.5 | 3 057.9 |
| 1E+8 | 56 769.0 | 11 154.0 | 11 505.0 | 12 004.0 | 9 680.0 |

It can be seen that the best time performance is for the Gauss ring points. The time performance for Halton points and for uniform distribution of points inside a rectangle is similar. Overall, we can say that for all tested distributions of input points, except uniform points in an ellipse, is the running time practically similar. This is expected behavior because most of the points are eliminated during the phase of polar division. Therefore, there are only a few points and nonempty cells of uniform grid for finding the maximum distance. The worst time performance was obtained for uniform points in an ellipse.

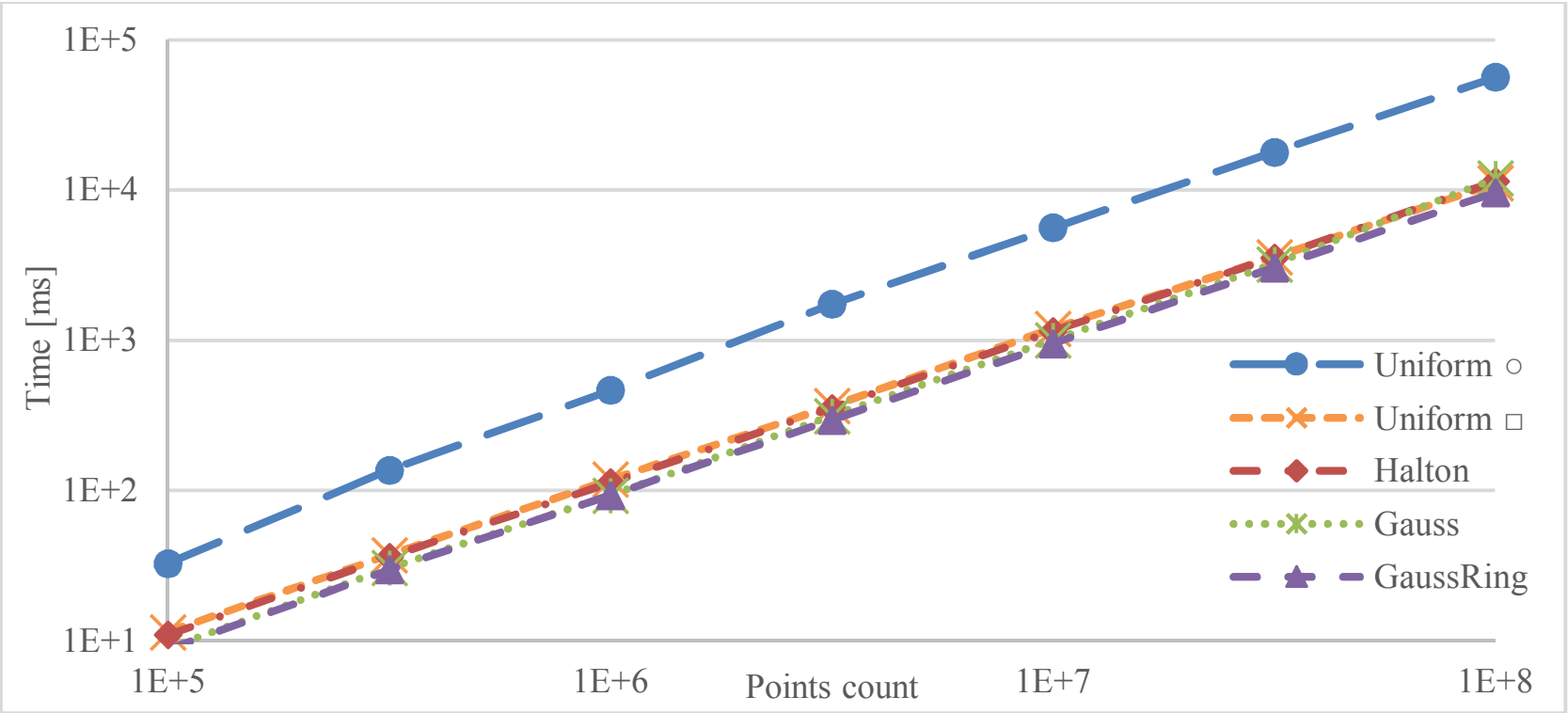

**Fig. 12.** The time performance of algorithm for finding maximum distance two points for different number of input points and different distribution of this points.

### 3.4 Comparison with Other Algorithms

We compared our proposed algorithm for finding exact maximum distance of two points in the given dataset with the BF algorithm, whose time complexity is $O(N^2)$, and with the algorithm proposed in [8], which has expected time complexity $O(N)$, where $N$ is the number of input points. It should be noted that the results for the algorithm in [8] are based on the use of the ratio of the BF algorithm to this algorithm.

Running times were measured for different numbers of input points with uniformly distributed points. The resultant speed-up of our proposed algorithm with respect to the BF algorithm and algorithm in [8] can be seen in Fig. 13 and Fig. 14.

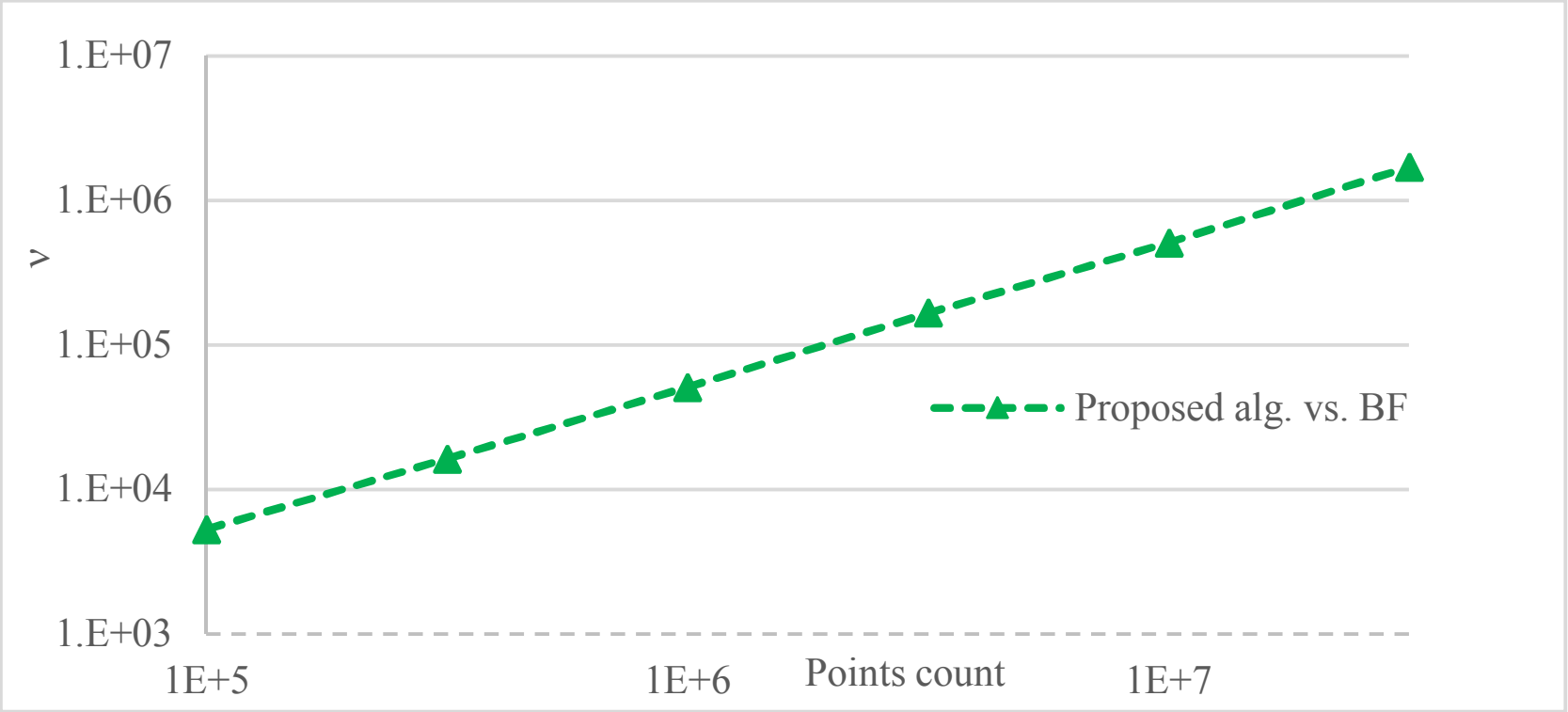

**Fig. 13.** The speed-up of our proposed algorithm for uniformly distributed points with respect to BF algorithm for the same datasets.

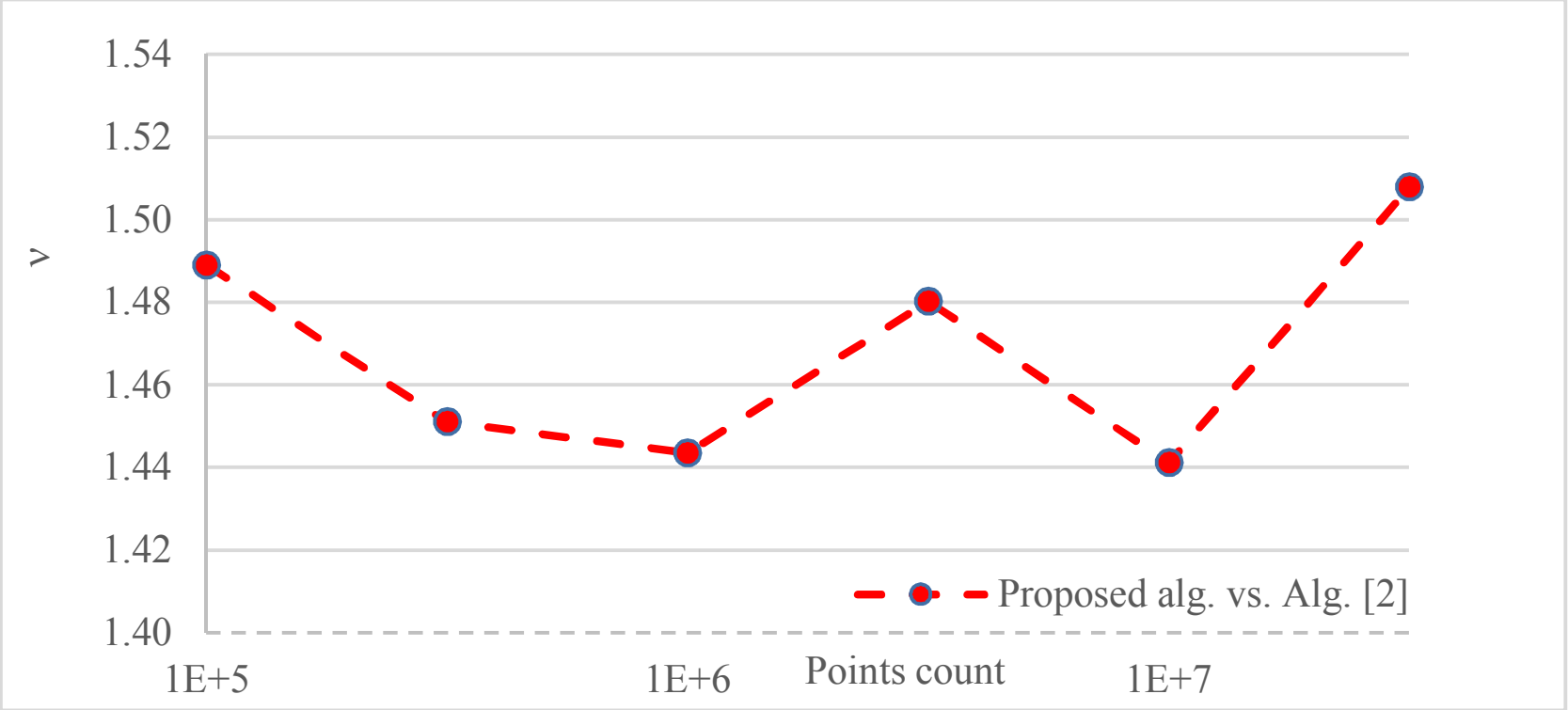

**Fig. 14.** The speed-up of our proposed algorithm for uniformly distributed points with respect to algorithm in [8] for the same datasets.

It can be seen that the speed-up of the proposed algorithm is significant with respect to BF algorithm and grows with the number of points processed. Moreover, our algorithm is in average 1.5 times faster than the algorithm in [8].

## 4 Conclusion

A new fast algorithm for finding an exact maximum distance of two points in $E^2$ with $O_{\text{expected}}(N)$ complexity has been presented. This algorithm uses a space division technique. It is robust and can process a large number of points. The advantages of our proposed algorithm are simple implementation and robustness. Moreover, our algorithm can be easily extended to $E^3$ by a simple modification.

For future work, the algorithm for finding exact maximum distance of two points, can be easily parallelized, as most of the steps are independent. The second thing is to extend this algorithm to $E^3$.

**Acknowledgments.** The authors would like to thank their colleagues at the University of West Bohemia, Plzen, for their discussions and suggestions, and anonymous reviewers for their valuable comments and hints provided. The research was supported by MSMT CR projects LH12181 and SGS 2013-029.

## References

1. Clarkson, K.L., Shor, P.W.: Applications of random sampling in computational geometry, II. Discrete & Computational Geometry, 1989, Vol.4, No.1, pp.387-421.
2. Dobkin, D. P., Snyder, L.: On a general method for maximizing and minimizing among certain geometric problems. Proceedings of the 20th Annual Symposium on the Foundations of Computer Science, 1979, pp. 9-17.
3. Fasshauer, G.E.: Meshfree Approximation Methods with MATLAB. World Scientific Publishing Co., Inc., 2007.
4. Hilyard, J., Teilhet, S.: C# cookbook. O'Reilly Media, Inc., 2006.
5. Liu, G., Chen, Ch.: A new algorithm for computing the convex hull of a planar point set, Journal of Zhejiang University SCIENCE A, 2007, Vol.8, No.8, pp.1210-1217.
6. Mehta, D. P., Sahni, S: Handbook of data structures and applications. CRC Press, 2004.
7. O'Rourke, J.: Computational geometry in C. Cambridge university press, 1998.
8. Skala, V.: Fast $O_{\text{expected}}$ (N) algorithm for finding exact maximum distance in E2 instead of O ($N^2$) or O (N lgN). AIP Conference Proceedings, 2013, No.1558, pp.2496-2499.
9. Snyder, W.E., Tang, D.A.: Finding the extrema of a region. IEEE Trans. on Pattern Analysis and Machine Intelligence, 1980, Vol.2, No.3, pp.266-269.
10. Vince, J.: Geometric algebra for computer graphics. Springer Science & Business Media, 2008.